\pdfoutput=1  
\def\cO{\cos\Omega_\mathrm{b}}
\def\sO{\sin\Omega_\mathrm{b}}
\def\cI{\cos I_\mathrm{b}}
\def\sI{\sin I_\mathrm{b}}
\def\nk{n_\mathrm{b}}

\def\kl{{\boldsymbol{\hat{J}}_\mathrm{b}}\boldsymbol{\cdot}{\boldsymbol{\hat{l}}}_\mathrm{b}}
\def\km{{\boldsymbol{\hat{J}}_\mathrm{b}}\boldsymbol{\cdot}{\boldsymbol{\hat{m}}}_\mathrm{b}}
\def\kh{{\boldsymbol{\hat{J}}_\mathrm{b}}\boldsymbol{\cdot}{\boldsymbol{\hat{h}}}_\mathrm{b}}
\def\Pb{P_\mathrm{b}}
\def\rfr#1{Equation\,(\ref{#1})}
\def\rfrs#1#2{Equations\,(\ref{#1})--(\ref{#2})}

\def\virg#1{``#1"}
\def\eqi{\begin{equation}}
\def\eqf{\end{equation}}
\def\rp#1#2{\frac{#1}{#2}}
\def\lb#1{\label{#1}}

\def\ton#1{\left(#1\right)}
\def\qua#1{\left[#1\right]}
\def\grf#1{\left\{#1\right\}}

\RequirePackage[2020-02-02]{latexrelease}
\documentclass{aastex631}
\usepackage{morefloats}
\usepackage[title]{appendix}
\usepackage{textcomp}
\usepackage{booktabs}
\usepackage{multirow}
\usepackage{rotating,tabularx}
\usepackage{float}
\usepackage{enumerate}
\usepackage{rotating}
\usepackage[polutonikogreek,english]{babel}
\usepackage{amsmath,starfont,textgreek,w-greek}
\usepackage[flushleft]{threeparttable}
\usepackage{amsthm}
\usepackage{amscd}
\usepackage[mathlines]{lineno}
\usepackage{amssymb,dsfont}
\usepackage{graphicx,epsfig}
\usepackage{txfonts}
\usepackage{xr-hyper}
\usepackage{hyperref}

\usepackage[nointegrals]{wasysym}
\usepackage[caption=false]{subfig}

\allowdisplaybreaks

\makeatletter
 \DeclareRobustCommand\ref{%
    \@ifstar\@refstar\T@ref
  }%
  \DeclareRobustCommand\pageref{%
    \@ifstar\@pagerefstar\T@pageref
  }%
 \makeatother

\begin{document}

\title{Measuring a gravitomagentic effect with the triple pulsar PSR J0337+1715}

\shortauthors{L. Iorio}

\author[0000-0003-4949-2694]{Lorenzo Iorio}
\affiliation{Ministero dell' Istruzione e del Merito. Viale Unit\`{a} di Italia 68, I-70125, Bari (BA),
Italy}

\email{lorenzo.iorio@libero.it}

\begin{abstract}
To the first post--Newtonian order, the orbital angular momentum of the fast--revolving inner binary of the triple system PSR J0337+1715, made of a millisecond pulsar and a white dwarf, induces an annular gravitomagnetic field which displaces the line of apsides of the slower orbit of the other, distant white dwarf by $-1.2$ milliarcseconds per year. The current accuracy in determining the periastron of the outer orbit is $63.9$ milliarcseconds after 1.38 years of data collection. By hypothesizing a constant rate of measurement of the pulsar's times of arrivals over the next 10 years, assumed equal to the present one, it can be argued that the periastron will be finally known to a $\simeq 0.15$ milliarcseconds level, while its cumulative gravitomagnetic retrograde shift will be as large as $-12$ milliarcseconds. The competing post--Newtonian gravitolectric periastron advance due to the inner binary's masses, nominally amounting to $74.3$ milliarcseconds per year, can be presently modelled to an accuracy level as good as $\simeq 0.04$  milliarcseconds per year. The mismodelling in the much larger Newtonian periastron rate due to the quadrupolar term of the multipolar expansion of the gravitational potential of a massive ring, whose nominal size for PSR J0337+1715  is $0.17$ degrees per year, might be reduced down to the $\simeq 0.5$ milliarcseconds per year level over the next 10 years. Thus, a first measurement of such a novel form of gravitomagnetism, although challenging, may be somehow feasible in a not too distant future.
\end{abstract}


\section{Introduction}
PSR J0337+1715 \citep{2014Natur.505..520R} is a peculiar hierarchical triple system made of three astrophysical compact objects: an inner, tight binary $\mathcal{S}$ composed by a $2.7$ ms millisecond pulsar  A  and a white dwarf B revolving one around each other in a relative circular orbit with orbital period $\Pb \simeq 1.6\,\mathrm{d}$, and another white dwarf C moving about $\mathcal{S}$ along a wider circular path with orbital period $P \simeq 327\,\mathrm{d}$ and coplanar with that of $\mathcal{S}$ itself, both inclined by\footnote{The orbital parameters of the outer companion are written without any label, while those of $\mathcal{S}$ are dubbed with \virg{b}.} $I_\mathrm{b} = I=39.2^\circ$ to the plane of the sky, assumed as reference $\grf{x,\,y}$ plane. Their masses are $M_\mathrm{A} = 1.44 M_\odot,\,M_\mathrm{B} = 0.2 M_\odot$ and $M_\mathrm{C} = 0.4 M_\odot$, respectively, where $M_\odot$ is the mass of the Sun.

So far, PSR J0337+1715 is the sole relatively tight member so far discovered of the class of hierarchical triple systems hosting stellar corpses, apart from B162-26 whose pulsar has a white dwarf as inner companion, and a roughly Jupiter-mass at 35 au as outer orbiter \citep{1999ApJ...523..763T}. According to \citet{2014Natur.505..520R}, $\lesssim 1\%$ of the millisecond pulsars population resides in stellar triples, and $\lesssim 100$ such systems exist in the Galaxy. In general, it is expected that, in triple systems hosting a millisecond pulsar, dubbed as triple pulsars in the following, the orbit of the distant companion is more eccentric due to dynamical interactions between the stars during stellar evolution \citep{2011ApJ...734...55P}. The coplanarity and circularity of the orbits of PSR J0337+1715, along with the fact that $P/\Pb\simeq 200$, imply that the current configuration is stable on long timescales, which favored its discovery. It is likely that the angular momentum vectors of the inner and outer orbits were torqued into near alignment in a phase, lasted about 1 Gyr, during which the outermost member evolved and transferred mass onto the inner binary, perhaps within a common envelope \citep{1997MNRAS.285..288L}.

On the one hand, PSR J0337+1715 proved unsuitable, at least until now, to perform the usual tests of the General Theory of Relativity (GTR) done with some tight binary pulsars \citep{2006Sci...314...97K,2020Univ....6..156W,2021PhRvX..11d1050K} like, e.g., the Hulse--Taylor pulsar PSR B1913+16 \citep{1975ApJ...195L..51H} and the double pulsar PSR J0737-3039A/B \citep{2003Natur.426..531B,2004Sci...303.1153L} because of its orbital configuration. Indeed, the argument of periastron $\omega$ is badly defined because of the almost vanishing eccentricity $e$ of the orbit of the outer white dwarf. Furthermore, the mass of the inner one is far too small for the gravitational redshift of the pulsar  signal to be measurable. Finally, the Shapiro delay is negligible; the radio waves traveling along the line of sight toward the Earth pass very distant from its companion because the orbital plane is far from being seen edge--on.  On the other hand, PSR J0337+1715 turned out a valuable laboratory to put the tightest constraints so far  on the (absence of the) Nordtvedt effect \citep{1968PhRv..170.1186N,1968PhRv..169.1014N}. The latter is an orbital consequence of a possible violation of the strong equivalence principle which would take place should bodies with different amounts of gravitational self-energy, just like  a neutron star and a white dwarf, fell with different accelerations in an external gravitational field. As a consequence, if the falling objects orbit one around each other while moving altogether about a third body, their barycentric orbits should experience a differential elongation toward the source of the external field. Actually, no Nordtvedt effect was found in PSR J0337+1715 to a relative accuracy of two parts per million at $95\%$ confidence level \citep{2018Natur.559...73A,2020A&A...638A..24V}.

Does PSR J0337+1715 offer the possibility of testing some other predictions of GTR in a not too far future? The answer may become positive within a reasonable timeframe, as soon as a sufficiently large number of times of arrivals (TOAs) have been collected.

The paper is organized as follows. Section \ref{ringo} is devoted to the orbital precessions of a test particle induced by the gravitomagnetic field of a distant,  inner  rotating massive ring. In Section \ref{pulZar}, they are calculated for PSR J0337+1715 (Section \ref{Lets}), and an evaluation of the experimental accuracy obtainable in a not too remote future is given (Section \ref{erro}). The competing effects due to the  Schwarzschild--like gravitoelectric field of $\mathcal{S}$ (Section \ref{Scva}), and to the expansion of the Newtonian gravitational field of a massive ring to the quadrupolar order (Section \ref{vazo}) are worked out as well. Section \ref{fine} summarizes our findings and offers conclusions.
\section{The gravitomagnetic orbital precessions due to the orbital angular momentum of the inner binary}\lb{ringo}
The effect in question, to the first post-Newtonian (1pN) order, would not only be the usual gravitoelectric periastron advance due solely to the masses of the system, but, in principle, also a particular form of gravitomagnetic orbital shift.

At first, this might sound strange, given that the generalizations  of the  well--known Lense-Thirring (LT) orbital precessions \citep{1918PhyZ...19..156L} to a binary system \citep{Sof89}
\begin{align}
\dot I^\mathrm{LT}_\mathrm{b} \lb{dIdtLT} & = \rp{2GJ_\mathrm{b}\ton{\kl}}{c^2 a_\mathrm{b}^3\ton{1 - e_\mathrm{b}^2}^{3/2}}, \\ \nonumber \\
\dot \Omega_\mathrm{b}^\mathrm{LT} \lb{dOdtLT} & = \rp{2GJ_\mathrm{b}\ton{\km}}{c^2 \sin I_\mathrm{b} a_\mathrm{b}^3\ton{1 - e_\mathrm{b}^2}^{3/2}}, \\ \nonumber \\
\dot \omega_\mathrm{b}^\mathrm{LT} \lb{dodtLT} & = -\rp{2GJ_\mathrm{b}\qua{2\ton{\kh}+\cot I_\mathrm{b}\ton{\km}}}{c^2 a_\mathrm{b}^3\ton{1 - e_\mathrm{b}^2}^{3/2}},
\end{align}
induced by the spin angular momenta $\boldsymbol{J}_\mathrm{A}$ and $\boldsymbol{J}_\mathrm{B}$ of both bodies through
\eqi
{\boldsymbol{J}}_\mathrm{b}:= \ton{1 + \rp{3}{4}\rp{M_\mathrm{B}}{M_\mathrm{A}}}{\boldsymbol{J}}_\mathrm{A} + \ton{1 + \rp{3}{4}\rp{M_\mathrm{A}}{M_\mathrm{B}}}{\boldsymbol{J}}_\mathrm{B},
\eqf
are notoriously orders of magnitude smaller than the 1pN gravitoelectric periastron rate
\eqi
\dot\omega_\mathrm{b}^\mathrm{GE} = \rp{3\nk\mu_\mathrm{b}}{c^2 a_\mathrm{b}\ton{1 - e_\mathrm{b}^2}}.\lb{oGE}
\eqf
In \rfrs{dIdtLT}{oGE}, $G$ is the Newtonian constant of gravitation, $\mu_\mathrm{b} := G\ton{M_\mathrm{A} + M_\mathrm{B}}$ is the standard gravitational parameter of the binary, $c$ is the speed of light in vacuum, $a_\mathrm{b}, e_\mathrm{b},\omega_\mathrm{b}$ and $\Omega_\mathrm{b}$  are the semimajor axis, the eccentricity, the argument of pericentre and the longitude of the ascending node of the binary's relative orbit, respectively, while $\nk:=\sqrt{\mu_\mathrm{b}/a_\mathrm{b}^3}$ is the Keplerian mean motion related to the orbital period by $\Pb = 2\pi/\nk$. Furthermore,
\begin{align}
{\boldsymbol{\hat{l}}}_\mathrm{b} \lb{elle}&:=\grf{\cO,~\sO,~0}, \\ \nonumber \\
{\boldsymbol{\hat{m}}}_\mathrm{b} \lb{emme}&:=\grf{-\cI\sO,~\cI\cO,~\sI},\\ \nonumber \\
{\boldsymbol{\hat{h}}}_\mathrm{b} \lb{hacca}& :=\grf{\sI\sO,~-\sI\cO,~\cI}
\end{align}
are three mutually orthogonal unit vectors directed along the line of nodes in the reference $\grf{x,\,y}$ plane, a direction in the orbital plane perpendicular to the latter, and  the orbital angular momentum, respectively, in such a way that ${\boldsymbol{\hat{l}}}_\mathrm{b}\boldsymbol{\times}{\boldsymbol{\hat{m}}}_\mathrm{b} = {\boldsymbol{\hat{h}}}_\mathrm{b}$.
Suffice it to say that, for, e.g., the double pulsar PSR J0737–3039A/B \citep{2003Natur.426..531B,2004Sci...303.1153L}, the gravitoelectric periastron precession of \rfr{oGE} is as large as \citep{2006Sci...314...97K}
\eqi
\dot\omega_\mathrm{b}^\mathrm{GE} = 16.89\,\mathrm{deg\,yr^{-1}},
\eqf
while the gravitomagnetic one, mainly due to the angular momentum of PSR J0737–3039A, is expected to be as small as \citep{2020MNRAS.497.3118H,2021MNRAS.507..421I}
\eqi
\dot\omega_\mathrm{b}^\mathrm{LT}\simeq -0.00060\,\mathrm{deg\,yr^{-1}} = -2.16\,\mathrm{arcsec\,yr^{-1}};\lb{oLTA}
\eqf
Intense efforts to measure it, not yet crowned with success, are currently underway \citep{Kehletal017,2020MNRAS.497.3118H,HuFre2024}. The figure quoted in \rfr{oLTA} is inferred from\footnote{The angular momentum of the B component of the double pulsar, which rotates more slowly, is smaller by about two orders of magnitude.}
\eqi
J_\mathrm{A}\simeq 4.5\times 10^{40}\,\mathrm{kg\,m^2\,s^{-1}}
\eqf
for the angular momentum of  PSR J0737–3039A; it can be calculated by assuming for the pulsar's moment of inertia (MOI) the value \citep{2021PhRvL.126r1101S}
\eqi
\mathcal{I}_\mathrm{A}\simeq 1.6\times 10^{38}\,\mathrm{kg\,m^2}
\eqf
and recalling that the spin period of the A component of the double pulsar is $22.7\,\mathrm{ms}$ \citep{2006Sci...314...97K}.

As pointed out in \citet{2022Univ....8..546I}, a hierarchical triple system offers, in principle, the possibility of measuring the orbital precessions due to the gravitomagnetic field generated by the mass current generated by the fast orbital motion of the inner binary, seen as a rotating matter ring. In other words, one has to replace\footnote{The gravitomagnetic field induced by a rotating matter ring is, far from the latter, identical to the usual one for a rotating body whose angular momentum $\boldsymbol{J}$ is replaced with the ring's one \citep{2016Ap&SS.361..140R}. } ${\boldsymbol{J}}_\mathrm{b}$ in \rfrs{dIdtLT}{dodtLT} with the orbital angular momentum
${\boldsymbol{L}}_\mathrm{b}$ of the inner binary. Its magnitude is
\eqi
L_\mathrm{b} = M_\mathrm{red}\sqrt{\mu_\mathrm{b}a_\mathrm{b}\ton{1 - e^2_\mathrm{b}}},
\eqf
where
\eqi
M_\mathrm{red}:=\rp{M_\mathrm{A}M_\mathrm{B}}{M_\mathrm{b}}
\eqf
is the reduced mass, and
\eqi
M_\mathrm{b}:= M_\mathrm{A} + M_\mathrm{B}
\eqf
is the total mass of the inner binary.
Usually, $L_\mathrm{b}$ can be orders of magnitude larger than the spin angular momentum of any of the inner binary's components; in the case of PSR J0337+1715, it is
\eqi
L_\mathrm{b} = 3.5\times 10^{44}\,\mathrm{kg\,m^2\,s^{-1}},
\eqf
while the pulsar's spin angular momentum $J$ should be about three orders of magnitude smaller. From \rfrs{dIdtLT}{dodtLT} it straightforwardly follows that, for a coplanar triple system, i.e., if ${\boldsymbol{\hat{L}}}_\mathrm{b}\boldsymbol{\cdot}\boldsymbol{\hat{h}}=1,
\,{\boldsymbol{\hat{L}}}_\mathrm{b}\boldsymbol{\cdot}\boldsymbol{\hat{l}}={\boldsymbol{\hat{L}}}_\mathrm{b}\boldsymbol{\cdot}\boldsymbol{\hat{m}}=0$, the inclination and the node stay constant, while the periastron precesses at a rate
\eqi
\dot\omega_\mathrm{LT} = -\rp{4GL_\mathrm{b}}{c^2 a^3\ton{1 - e^2}^{3/2}}.\lb{rumba}
\eqf
\section{The case of PSR J0337+1715}\lb{pulZar}
\subsection{The effect of the 1pN gravitomagnetic annular field}\lb{Lets}
In the particular case of PSR J0337+1715,  \rfr{rumba} predicts for it a gravitomagnetic precession of the order of
\eqi
\dot\omega_\mathrm{LT} = -1.2\,\mathrm{mas\,yr^{-1}},
\eqf
where mas stands for milliarcseconds, while the current level of uncertainty in measuring its periastron over $1.38$ yr during which
\eqi
N_0 = 26280\lb{TOA}
\eqf
TOAs were collected, can be calculated from Table 1 of \citet{2014Natur.505..520R} to be
\eqi
\sigma^0_\omega\simeq 63.9\,\mathrm{mas}.\lb{erro}
\eqf
\subsection{Prospects for future accuracy improvements}\lb{erro}
By provisionally assuming that about same number of TOAs as given by \rfr{TOA} will be collected in $1.38$ yr over, say, the next 10 years, the resulting accuracy would be improved down to\footnote{Such an estimate is obtained by dividing \rfr{erro} by $\sqrt{N}$, where $N=\ton{10/1.38}\times N_0 = 190435$.} \eqi\sigma_\omega\simeq 0.15\,\mathrm{mas}\eqf
with respect to the figure quoted in \rfr{erro}, while the total gravitomagnetic shift would amount to
\eqi
\Delta\omega_\mathrm{LT} \simeq -12\,\mathrm{mas}.
\eqf
\subsection{The 1pN gravitoelectric periastron precession}\lb{Scva}
The 1pN gravitoelectric apsidal rate of the orbit of the outer companion around the inner binary, whose nominal value is
\eqi
\dot\omega_\mathrm{GE} = 74.3\,\mathrm{mas\,yr}^{-1},
\eqf
could be safely subtracted from the measured total periastron precession since its uncertainty, calculated by propagating in \rfr{oGE} the errors of $P,\mu,e$ from Table 1 of \citet{2014Natur.505..520R},  is currently as little as
\eqi
\sigma_{\dot\omega_\mathrm{GE}} = 0.04\,\mathrm{mas\,yr}^{-1}.
\eqf
\subsection{The Newtonian periastron precession due to the quadrupolar term of the gravitational potential of a matter ring}\lb{vazo}
By expanding  the Newtonian gravitational potential of a circular massive ring, calculated in points far from the latter and lying in its plane \citep{2009EJPh...30..623C},  to the quadrupolar order, an extra--potential falling as $1/r^3$ arises; in the case of the inner binary of PSR J0337+1715, it can be cast into the form \citep{2006physics...2034D}
\eqi
U_\mathrm{qp} = -\rp{\mu_\mathrm{b} a_\mathrm{b}^2}{4r^3}\lb{Uqp},
\eqf
where the ring's mass and radius are assumed to be equal to the total mass $M_\mathrm{b}$ and to the semimajor axis $a_\mathrm{b}$ of $\mathcal{S}$, respectively.

It can be straightforwardly calculated that the periastron of the distant companion is shifted by \rfr{Uqp} according to \citep{2007PhRvD..75h2001A,2012AnP...524..371I}
\eqi
\dot\omega^\mathrm{qp} = \rp{3\mu_\mathrm{b}a^2_\mathrm{b}}{4\sqrt{\mu a^7}\ton{1 - e^2}^2},\lb{dodt}
\eqf
where
\eqi
\mu:=G\ton{M_\mathrm{A} + M_\mathrm{B} + M_\mathrm{C}}
\eqf
is the total standard gravitational parameter of PSR J0337+1715.
The nominal precession of \rfr{dodt} amounts to
\eqi
\dot\omega^\mathrm{qp} = 636775\,\mathrm{mas\,yr}^{-1} = 0.17\,\mathrm{deg\,yr}^{-1}.\lb{kazzo}
\eqf

It turns out that the main sources of uncertainty in \rfr{dodt} are the masses of the pulsar A and of the outer white dwarf C.
By assuming that their errors will reduce by a factor of $1/\sqrt{N}$ in the next 10 yr, where $N$ is calculated as in Section \ref{erro}, the mismodeling in \rfr{kazzo} should finally be
\eqi
\sigma_{\dot\omega^\mathrm{qp}}\simeq 0.5\,\mathrm{mas\,yr}^{-1}.
\eqf
\section{Summary and conclusions}\lb{fine}
By assuming a rate of collection of TOAs over the next 10 years equal to the present one, it can be guessed that the accuracy in determining the periastron precession of PSR J0337+1715 may reach the $\simeq 0.15\,\mathrm{mas}$ level, allowing, in principle, for a measurement of its cumulative gravitomagnetic shift of $-12\,\mathrm{mas}$ due to the orbital angular momentum of the inner binary. The competing gravitoelectric advance should be subtracted from the total periastron variation without problems since its uncertainty is as of now as little as $0.04\,\mathrm{mas\,yr}^{-1}$. On the other hand, relying upon the aforementioned assumptions, it can be argued that the accuracy in modelling the nominally much larger Newtonian periastron precession due to the quadrupolar term of the multipolar expansion of the gravitational potential of a massive ring should reach the $\simeq 0.5\,\mathrm{mas\,yr}^{-1}$ level.

The situation may be more favorable for a hypothetical triple system, yet to be discovered, whose outer companion's orbit had a smaller size and was more eccentric, a scenario that should not be deemed as unrealistic. Furthermore, if ${\boldsymbol{\hat{L}}}_\mathrm{b}$ and $\boldsymbol{\hat{h}}$ were misaligned, also the inclination and the node precessions would come into play, as per \rfrs{dIdtLT}{dOdtLT}.
\section*{Data availability}
No new data were generated or analysed in support of this research.
\section*{Conflict of interest statement}
I declare no conflicts of interest.
\bibliography{Megabib}{}
\end{document}